\documentclass[sigconf]{acmart}

\usepackage{amsmath}
\usepackage{booktabs}
\usepackage{graphicx}
\usepackage{multirow}
\usepackage{xurl}
\usepackage{xcolor}
\usepackage{comment}
\usepackage{wasysym}
\usepackage{bm}
\usepackage{tcolorbox} 
\tcbuselibrary{listings,breakable}

\usepackage{setspace}
\usepackage{array}
\usepackage{tabularx}
\usepackage{calc}
\usepackage{makecell}
\usepackage{subfigure}
\usepackage{appendix}
\usepackage{enumitem}
\usepackage{color}
\usepackage{soul}
\usepackage{balance}
\usepackage{color, colortbl}

\newcommand{\CC}[1]{\cellcolor{gray!#1}}

\newcommand{\new}[1]{#1}

\renewcommand\footnotetextcopyrightpermission[1]{} 
\usepackage{pifont}
\newcommand{\cmark}{\ding{51}}%
\newcommand{\xmark}{\ding{55}}%

\usepackage{tikz}

\newcounter{criteria_counter}
\newcounter{rq_counter}

\AtBeginDocument{%
  \providecommand\BibTeX{{%
    \normalfont B\kern-0.5em{\scshape i\kern-0.25em b}\kern-0.8em\TeX}}}

\begin{document}

\title{On The Vulnerability of Anti-Malware Solutions to DNS Attacks}

\author{Asaf Nadler}
\affiliation{%
   \institution{Akamai Technologies and Ben-Gurion University of the Negev}
   \country{Israel}}
\email{asafnadl@post.bgu.ac.il}

\author{Ron Bitton}
\affiliation{%
\institution{Ben-Gurion University of the Negev}
\country{Israel}}
\email{ronbit@post.bgu.ac.il}

\author{Oleg Brodt}
\affiliation{%
\institution{Ben-Gurion University of the Negev}
\country{Israel}}
\email{bolegb@bgu.ac.il}

\author{Asaf Shabtai}
\affiliation{%
\institution{Ben-Gurion University of the Negev}
\country{Israel}}
\email{shabtaia@bgu.ac.il}

\begin{abstract}
Anti-malware agents typically communicate with their remote services to share information about suspicious files.
These remote services use their up-to-date information and global context (view) to help classify the files and instruct their agents to take a predetermined action (e.g., delete or quarantine).
In this study, we provide a security analysis of a specific form of communication between anti-malware agents and their services, which takes place entirely over the insecure DNS protocol.
These services, which we denote \emph{DNS anti-malware list (DNSAML)} services, affect the classification of files scanned by anti-malware agents, therefore potentially putting their consumers at risk due to known integrity and confidentiality flaws of the DNS protocol.

By analyzing a large-scale DNS traffic dataset made available to the authors by a well-known CDN provider, we identify anti-malware solutions that seem to make use of DNSAML services.
We found that these solutions, deployed on almost three million machines worldwide, exchange hundreds of millions of DNS requests daily.
These requests are carrying sensitive file scan information, oftentimes - as we demonstrate - without any additional safeguards to compensate for the insecurities of the DNS protocol.
As a result, these anti-malware solutions that use DNSAML are made vulnerable to DNS attacks.
For instance, an attacker capable of tampering with DNS queries, gains the ability to alter the classification of scanned files, without presence on the scanning machine.

We showcase three attacks applicable to at least three anti-malware solutions that could result in the disclosure of sensitive information and improper behavior of the anti-malware agent, such as ignoring detected threats. 
Finally, we propose and review a set of countermeasures for anti-malware solution providers to prevent the attacks stemming from the use of DNSAML services.

\end{abstract}


\maketitle
\pagestyle{plain}
\renewcommand{\shortauthors}{Nadler et al.}

\section{\label{sec:intro}Introduction}

Anti-malware agents are a popular security solution, running on millions of endpoints and servers on a regular basis to check files for malicious code.
Despite their popularity, anti-malware agents are fairly limited, because they lack a global context of the threat landscape and are not always updated with the most recent threats.
For these reasons, when encountering a file that may carry an unknown threat, an anti-malware agent will typically send information related to the inspected file to a remote service (e.g.,~\cite{oberheide2008cloudav, wiki:cloudav}).
Such services integrate information from multiple agents worldwide for improved protection against newly emerging threats. 
Based on the information it receives, the service classifies the file and instructs its agents to perform predetermined actions (e.g., quarantine or delete the suspicious file) based on this classification.

The transfer of information regarding scanned files between anti-malware agents and their services, and the files' classification must be secured to ensure integrity and confidentiality.
Otherwise, attackers may be able to learn which files are located on the endpoint, whether they contain malicious code, and far worse --- interfere with the agent's actions regarding the code/file inspected.
This raises a concern regarding anti-malware solutions that are based on agents and services that utilize the insecure Domain Name System (DNS) protocol for the information delivery~\cite{mcafee_gti,sophos_sxl,cymru_mhr,eset_livegrid}.
In this case, we refer to the remote services of an anti-malware solution as \emph{DNS anti-malware list (DNSAML) services}.

\begin{figure} [t]
	\centering
	\includegraphics[width=0.47\textwidth]{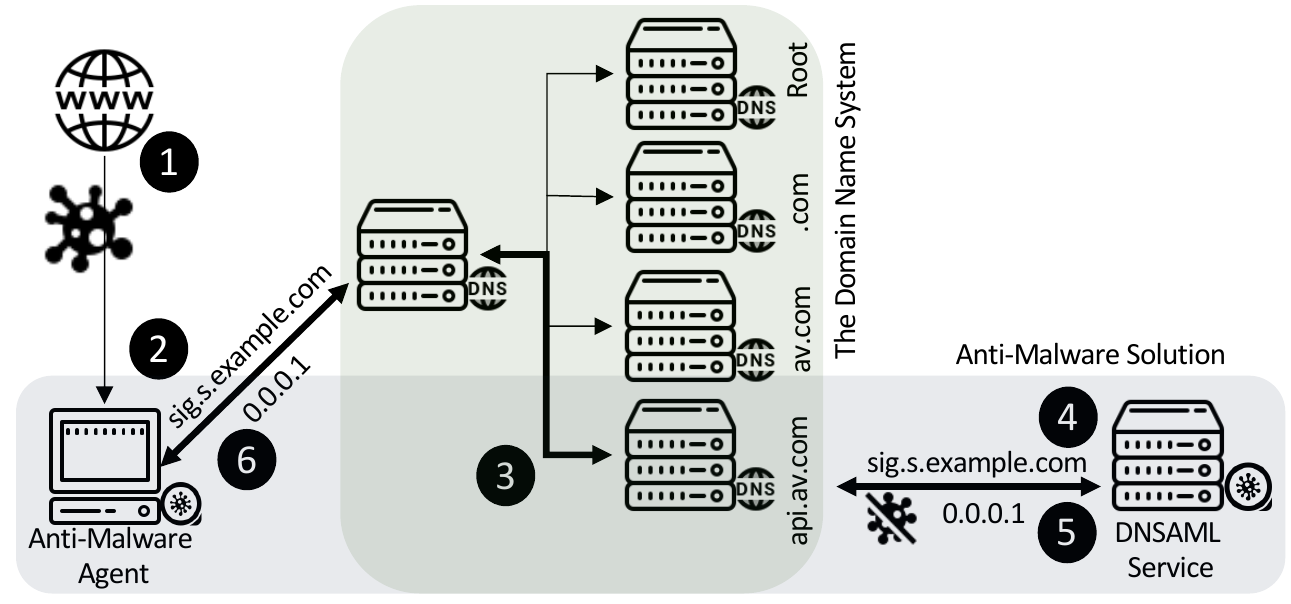}
	\caption{An illustration of an anti-malware agent query to its DNSAML service: an anti-malware agent that downloads a malicious file (1) scans the file and reports its signature to its DNSAML service (2-4), and based on the service response (5), the agent performs a predetermined action such as deleting the file (6).}
	\label{fig:scheme}
\end{figure}

DNSAML services are queried by anti-malware agents, as illustrated in Figure~\ref{fig:scheme}:
When an endpoint (or server) protected by an anti-malware agent faces a suspicious file, it matches the file's signature against the local database that contains signatures of known malware (step 1). 
If the agent fails to find a match, it issues a DNS query, prepended with a malware file signature (e.g., hash) to a DNS zone owned by the global threat intelligence service (step 2).
The DNS query, transmitted through the domain name system (step 3), is eventually directed to the DNSAML service (step 4).
The service matches the signature prepended to the DNS query against its up-to-date database and issues a DNS response back to the agent indicating whether the file is malicious (step 5).
Finally, based on the response, the agent learns whether the scanned file is malicious and acts in a predefined manner (e.g., quarantining or deleting the file, issuing an alert, or taking no action) (step 6).

The DNSAML service architecture is heavily inspired by that of Domain Name System Block Lists (DNSBLs)~\cite{wiki:dnsbl}, an architecture established to assist website administrators with blocking incoming connections from systems associated with spam.
The similar architecture provides DNSAML services with advantages like those of DNSBLs, such as ease-of-use and fast performance.
However, this architecture has disadvantages from the security perspective, because the DNS protocol is known to suffer from privacy and integrity flaws, such as a lack of source authentication and data encryption. 
These flaws enable a man-in-the-middle attacker to read, manipulate, and tamper with DNS packets without proper verification by the DNS protocol and its components~\cite{yan2020road,deccio2019dns}.
Therefore, anti-malware solutions that use DNSAML services without implementing additional safeguards may expose their solution and their consumers to various attacks that can result in information disclosure and affect decision-making.
For example, an attacker capable of manipulating DNS queries may respond on behalf of the DNSAML service and provide a false classification in order to incorrectly convince an anti-malware agent that a benign scanned file is malicious, or vice versa, causing the agent to delete benign files or maintain malicious files, respectively.

\subsection{Research Questions}
In this study we perform a security analysis regarding the use of DNSAML services by anti-malware solutions.
Specifically, we attempt to answer the following research questions in Sections~\ref{sec:usingdns}-\ref{sec:countermeasures}:

\begin{list}{\textbf{RQ \arabic{rq_counter}.}}
{
\usecounter{rq_counter}
\setlength\labelwidth{0.4in}
\setlength\leftmargin{0.4in}
\setcounter{rq_counter}{0}
}
    \item Which anti-malware solutions make use of DNSAML services and how prevalent are these solutions?
    \item What threats stem from the use of DNSAML services?
    \item How secure are anti-malware solutions in light of the identified threats?
    \item Which countermeasures can prevent attacks stemming from the insecure use of DNSAML services on anti-malware solutions and their consumers?
\end{list}
\subsection{Methodology}
To identify DNSAML services and examine their prevalence, we perform an analysis on a proprietary, large-scale DNS traffic dataset that was shared with the authors by one the world's largest CDN providers (see Section~\ref{sec:usingdns}).
This dataset includes 30 days of DNS traffic, with an average of \emph{52 billion DNS queries daily} to more than 300 million registered domain names by 37 million Internet machines scattered across 29 countries.

We identify DNSAML services by searching for DNS zones that are (a) widely queried with (b) prepended payloads of structured information, such as binary hash signatures, and (c) categorized by web categorization services as security-related.
This search results in 55 services, from which, we select five specific DNSAML services that are well-documented to support our analysis.

For the selected DNSAML services, we examine the DNS queries made to their corresponding DNS zones.
The results of the analysis show that DNSAML services process more than 108 million queries daily which are made by at least 2.85 million worldwide anti-malware agents.
The anti-malware agents are installed on both endpoints and network gateways thus implying that end-users are exposed to potential threats both directly by their endpoint and indirectly by their network gateways.
These results emphasize the importance of a threat analysis of the potentially insecure use of the DNS protocol by these applications. 

To perform the threat analysis, we define a threat model where attackers with DNS eavesdropping and/or data tampering capabilities attempt to carry out three practical attacks:
\begin{enumerate}[leftmargin=*]
    \item an attacker with DNS eavesdropping capability that can learn what files were downloaded by endpoints, whether endpoints have downloaded malware, and which IP addresses they communicate with, thus indicating potential information disclosure;
    \item an attacker with DNS data tampering capability that can convince anti-malware agents that malicious files are benign and thereby prevent their deletion;
    \item an attacker with DNS data tampering capability that can convince anti-malware agents that benign files are malicious, resulting in false alerts, quarantines, or deletion of benign files.
\end{enumerate}

To demonstrate the feasibility of the attacks, we have constructed a setup in which a victim machine downloads a file and performs a scan, and an attacker machine that intercepts and spoofs the DNS responses of the DNSAML service successfully gains information about the scanned file and is able to alter its classification by the malware agent.
This demonstration is conducted separately on three anti-malware agents (their most recent versions), thus showcasing the applicability of the attacks. (see Section~\ref{sec:in-depth})

\subsection{Main Findings}
Our findings show that under the defined threat model, three attacks are applicable to at least three well-known anti-malware solutions (see Section~\ref{sec:in-depth}).
Based on our analysis, agents of these vulnerable anti-malware solutions are installed on more than 2.6M endpoints and servers, and are used to perform over 108 million file scans every day.
Accordingly, the consumers of these anti-malware solutions are at risk.
Their security is compromised, due to the failure of their anti-malware solution to secure their file scan information, leaving them vulnerable to information disclosure and putting their files at risk of misclassification.

We submitted a responsible disclosure and shared our results with the anti-malware solution providers (Appendix~\ref{sec:responsible_disclosure}), and these providers are making changes to improve the security of their solutions. 
We also propose a set of countermeasures for anti-malware solution providers to help them prevent the attacks ( Section~\ref{sec:countermeasures}).

\subsection{Summary of Contributions}
The contributions of this study are as follows:
\begin{enumerate}[leftmargin=*]
    \item We identify the need for a security analysis of DNSAML services used by anti-malware solutions;
    \item We present a method for searching and identifying DNSAML services that can be used in future threat analysis;
    \item We define a threat model under which three attacks made possible due to the use of DNSAML services are shown to be applicable to at least three well-known anti-malware solutions;
    \item We propose and review a set of countermeasures for anti-malware solution providers to prevent the attacks.
\end{enumerate}

\section{\label{sec:usingdns}DNS Anti-Malware List Services}

In this section, we analyze a large-scale dataset of real DNS traffic in order to: (a) identify anti-malware solutions that make use of DNSAML services, and (b) evaluate the extent of their use.

\subsection{\label{subsec:dataset}Dataset}
We base our analysis on DNS traffic observed on a private, large-scale dataset that was shared with the authors by a well-known CDN provider. 
The dataset includes 30 days of DNS traffic logs recorded between November 1-30, 2020.
\new{The logs were recorded by recursive DNS servers that support \emph{only} plain-text DNS, thus ensuring that the analyzed traffic was never encrypted with either DNS-over-HTTPS or DNS-over-TLS.}

For each day in the dataset, there are 52 billion DNS queries, on average, to more than 300 million registered domain names, performed by at least 37 million machines scattered across 29 worldwide Internet service providers (ISPs).
\new{28 out of 29 ISPs are local ISPs that provide DNS services within a single, specific country, and all together cover 20 countries.
The remaining ISP is worldwide and covers various and indistinguishable countries.}
Every record within the dataset describes a single DNS query made, its response, \new{the ISP providing the DNS services, and an obfuscated string that matches the originating IP address of the querying machine that ensures privacy and anonymity}.

\subsection{\label{subsec:identify}Identification of DNSAML Services}
The DNSAML services used by anti-malware solutions are expected to meet the following criteria:\\
\noindent \textbf{Active (C1)}: DNSAML services must process a high volume of DNS queries. \\
\noindent \textbf{Security-related (C2)}: The DNS zones (i.e., domain names) that are used to query the DNSAML services must be registered and owned by an anti-malware solution provider. \\
\noindent \textbf{Structured DNS communication (C3)}: DNS queries made to the service include IP address, binary hash signatures, and/or encrypted telemetries in various forms. 

Based on these criteria, we propose the following method to search for registered DNS zones suspected as DNSAML services:
\begin{tcolorbox}[left=0pt,breakable,title=\textit{DNSAML service search}]
\small
    \begin{enumerate}\leftmargin=1em
        \item \textbf{C1}: Collect $X_{C1}$: a distinct list of registered DNS zones that are queried at least $\theta$ times a day, on average, i.e., active domains
        \item \textbf{C2}: Collect $X_{C2}$: a list of domain names owned by security services, including anti-malware solution providers.
        \item \textbf{C3}: Collect $X_{C3}$: a list of registered DNS zones, whose DNS queries facilitate structured messages in $X_{C1} \cap X_{C2}$ as follows:
        \begin{enumerate}
            \item Create $S_{IP}$: the set of zones in $X_{C1} \cap X_{C2}$ for which at least 10\% of their DNS queries consist of a valid IPv4 address.
            \item Create $S_{HASH}$, the set of zones in $X_{C1} \cap X_{C2}$ for which at least 10\% of their DNS queries consist of a 32 character alphanumeric label.
            \item Create $S_{TUN}$ the set of zones in $X_{C1} \cap X_{C2}$ identified by a DNS tunneling classifier.
            \item $X_{C3} = S_{IP} \cup S_{HASH} \cup S_{TUN}$
        \end{enumerate}
        
        \item Return $X_{C1} \cap X_{C2} \cap X_{C3}$: a list of registered DNS zones satisfying C1, C2, and C3 and therefore suspected as DNSAML services.
    \end{enumerate}
\end{tcolorbox}

\begin{figure*}[ht]
    \centering
    \includegraphics[width=\textwidth]{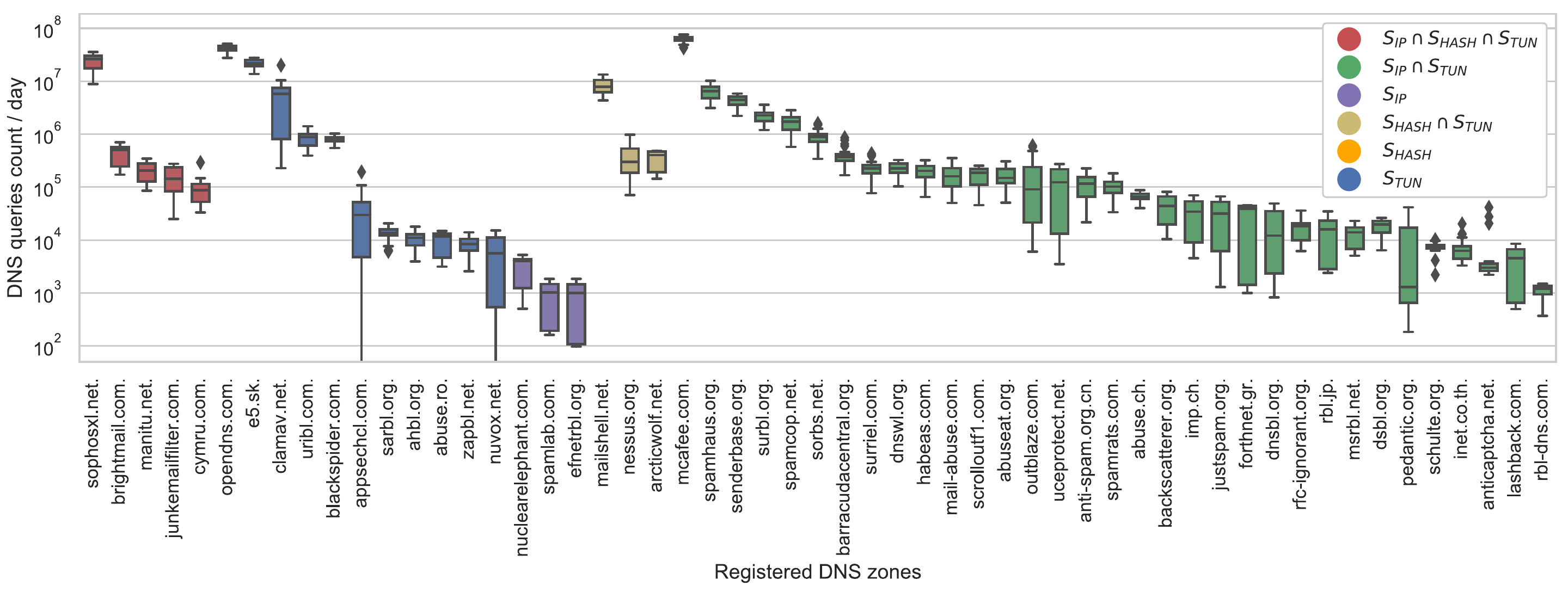}
    \caption{DNS queries to suspected DNSAML services for each identification method.}
    \label{fig:identifying}
\end{figure*}

The set of active registered DNS zones $X_{C1}$ included 184,046 registered domain names that were queried at least $\theta=1000$ times per day, on average (i.e., $0.6\%$ of the registered zones in the dataset).
The set of security-related DNS zones $X_{C2}$ was established using Webroot BrightCloud, one of the world's largest web classification services.
Specifically, we extracted the ``Computer and Internet Security'' category, resulting in 220,426 domain names.
The intersection of active and security-related domains $X_{C1} \cap X_{C2}$ included $4,884$ registered DNS zones.
The set of DNS zones whose queries facilitate structured messages $X_{C3}$, consisting of less than $0.017\%$ of the registered DNS zones in the dataset, was extracted from $X_{C1} \cap X_{C2}$ for efficiency purposes.
The set $X_{C3}$ included 55 registered DNS zones, of which 43 consisted of IPv4 addresses ($S_{IP}$); ten consisted of 32 bit hashes ($S_{HASH}$), and 51 were identified by a DNS tunneling solution ($S_{TUN}$) proposed by Nadler et al.~\cite{nadler2019detection} to classify domain names related to DNS tunneling traffic.
The set of suspected DNSAML services $X_{C1} \cap X_{C2} \cap X_{C3}$ and their activity appears in Figure~\ref{fig:identifying}, with their respective DNS queries per day (i.e., activity) and identification method, $S_{TUN}, S_{IP}, S_{HASH}$.

Finally, from the set of 55 suspected DNSAML services, we selected five services that we can strongly argue to be DNSAML services based on publicly available documentation:  

\noindent\textbf{Sophos' Extensible List (SXL)~\cite{sophos_sxl}}
The Sophos eXtensible List is a DNSAML service providing a malware hash lookup, IP reputation lookups, and web categorization.
The service is queried mainly by the Sophos Endpoint Cloud, a lean agent for endpoint devices that checks whether incoming emails are related spam and contain malware by verifying the reputation of the email sender, perform malware hash lookups for attachments, etc.

\noindent\textbf{McAfee's Global Threat Intelligence (GTI)~\cite{mcafee_gti}} 
McAfee's GTI is a DNSAML service that provides up-to-date malware detection for a number of Windows-based McAfee antivirus agents.
The list of possible agents includes McAfee's Endpoint Security, VirusScan Enterprise, and SaaS Endpoint Protection. 
These agents look for suspicious programs, Portable Document Format (PDF) files, and Android Package files and send a DNS request to the central GTI database hosted by McAfee Labs, in order to determine whether the file is malicious. 

\noindent\textbf{ESET's LiveGrid~\cite{eset_livegrid}}
ESET LiveGrid is a DNSAML service used to collect information about suspected files. 
The collected information is analyzed and processed by ESET malware experts.
Eventually, the experts update ESET detection engines with files that were classified as malicious in order to provide a faster reaction to malware and increase awareness of emerging threats.
The service is queried by ESET's ThreatSense endpoint agent

\noindent\textbf{Tenable's MalwareDB~\cite{wiki:nessus}}
Tenable's MalwareDB is a DNSAML service, which is queried mainly by Nessus Professional, an agent designed for vulnerability and malware scanning.

\noindent\textbf{Team Cymru's Malware Hash Registry (MHR)~\cite{cymru_mhr}}
Team Cymru's Malware Hash Registry (MHR) is a DNSAML service that aggregates the results of over 30 antivirus tools to assist in identifying unknown or suspicious files.
The service does not come with an official agent, however it is associated with informal open-source agents.
The service is queried more than 89,000 time per day, on average, (see Table~\ref{tab:dns-apis}) thereby suggesting that the service is used extensively for classifying malicious files.

\subsection{\label{subsec:prevalence}The Prevalence of DNSAML Services}

Based on the identified DNSAML services (presented in Section~\ref{subsec:identify}), we evaluate the extent of their use by analyzing the dataset described in Section~\ref{subsec:dataset}.
The extent of their use is evaluated using three measures: the number of daily file scans for classification delivered to the service, number of anti-malware agents communicating with the service on a daily basis, and number of countries from which the agents operate.

\noindent\textbf{Number of file classifications}
The mean number of daily DNS queries to DNSAML services is 108 million, each of which is assumed to include a binary file hash for classification.
This amount accounts for almost 0.2\% of the DNS queries within our large-scale DNS dataset.
Our interpretation of this is that the number of file scans performed through DNSAML services by anti-malware solutions indicates wide use meriting a thorough threat analysis.

\noindent\textbf{Number of agents}
The mean number of originating IP addresses that sent at least one DNS query to a DNSAML service is 2.85 million, each of which is assumed to be an anti-malware agent performing a file classification.
Our interpretation of the results is that a few million agents, as well as the machines on which they are hosted and the consumers they serve, are subject to the risks introduced by the use of DNSAML services.

\noindent Furthermore, the varying number of agents associated with the different DNSAML services leads us to argue that anti-malware solutions whose services are accessed by a large number of agents, e.g., above 10,000, are likely to involve endpoint installations, whereas anti-malware solutions with fewer than that might be used mainly on network gateways or endpoints.
Based on that, we argue that the DNSAML services associated with Sophos, McAfee, and ESET involve queries from endpoint installations.
\new{The rest of the selected DNSAML services, those associated with Tenable, and Team Cymru, are either deployed on a small number of endpoints and/or are queried primarily by a small number of websites and/or network gateways.}
This argument matches the partially available documentation of these anti-malware solutions (provided in Section~\ref{subsec:identify}) and leads us to conclude that users are exposed to potential risks, posed both directly by their machine through their endpoint agent and indirectly by their network servers, gateways, and proxies.

\noindent\textbf{Number of originating countries}
Anti-malware solutions that are more prevalent geographically imply a larger attack surface for man-in-the-middle attacks.
Within our dataset, there were 29 countries that we managed to identify with high accuracy. 
Based on our set of selected DNSAML services, the top scanning agents associated with McAfee, Sophos, and ESET perform scans from at least 24 countries in the dataset on a daily basis.
The other three agents perform scans from no less than 11 countries in the dataset on a daily basis.
We conclude that all of the examined agents operate globally, with endpoint-related agents being more globally spread than server-based agents.

\begin{table*}[ht]
\centering
\resizebox{\textwidth}{!}{%
\begin{tabular}{@{}lllllll@{}}
\Xhline{3\arrayrulewidth}
Company &
  \begin{tabular}[c]{@{}l@{}}Anti-Malware \\ Solution\end{tabular} &
  Agent &
  Zones &
  \begin{tabular}[c]{@{}l@{}}\# Daily File \\  Classifications (Mean) \end{tabular} &
  \begin{tabular}[c]{@{}l@{}}\# Daily Unique \\ Agents (Mean) \end{tabular} &
  \begin{tabular}[c]{@{}l@{}}\# Daily Agent \\ Origin Countries (Mean)\end{tabular} \\ \Xhline{3\arrayrulewidth}
Sophos     & SXL {[}4{]}        & Sophos Endpoint Cloud         & \url{*.sophosxl.net}             & 24078670.29	 & 80565	 &  24.74
\\
McAfee &
  GTI {[}1{]} &
  \begin{tabular}[c]{@{}l@{}}McAfee Endpoint Security\\ McAfee VirusScan Enterprise\\ McAfee SaaS Endpoint Protection\end{tabular} &
  \begin{tabular}[c]{@{}l@{}}\url{*.avts.mcafee.com}\\ \url{*.avqs.mcafee.com}\end{tabular} &
  62106752.81 & 2602658.81	
   & 28.77
   \\
ESET       & LiveGrid {[}6{]}   & ESET ThreatSense              & \url{*.e5.sk}            &      21807771.16                       & 164868.68 &  26.58 \\
Tenable    & MalwareDB {[}35{]} & Nessus Professional           & \url{*.l2.nessus.org}            &              374677.32               & 618.00  & 20.74 \\
Team Cymru & MHR {[}3{]}        & Cymru Services (Unofficial) & \url{*.malware.hash.cymru.com}   &          89802.48                   & 1822.00	 & 21.42\\ \Xhline{3\arrayrulewidth}
Total       &  &    &                              & $1.08 \cdot 10^{8}$ & $2.85 \cdot 10^{6}$ &  \\ \Xhline{3\arrayrulewidth}
\end{tabular}%
}
\caption{The mean number of files whose information is delivered to DNSAML services daily for classification, the number of agents communicating with DNSAML services, and the number of originating countries from which agents communicating with DNSAML services operate.}
\label{tab:dns-apis}
\end{table*}

\section{Threat Analysis\label{sec:threat_analysis}}
In this section, we define the threats introduced to anti-malware solutions and their consumers as a result of the insecure use of DNSAML services.

\subsection{Threat Model\label{subsec:threat_model}}

In our threat model, we distinguish between the following general attacker capabilities:
\begin{enumerate}[leftmargin=*]
    \item \textbf{DNS Eavesdropping Capability (CE):} An attacker that can \textit{inspect} \textit{all} DNS communication between the DNS resolver and forwarder.
    
    \item \textbf{Full DNS Tampering Capability (CT):} An attacker that can \textit{tamper} with \textit{all} DNS communication between the DNS resolver and forwarder.
    
    \item \textbf{Limited DNS Tampering Capability (LT):} An attacker that can \textit{tamper} with \textit{specific} DNS communication between the DNS resolver and forwarder.
\end{enumerate}

\newcolumntype{?}{!{\vrule width 1.5pt}}
\begin{table*}[t]
\small
\centering
\begin{tabular}{p{0.18\textwidth}|p{0.47\textwidth}?p{0.018\textwidth}p{0.018\textwidth}p{0.018\textwidth}p{0.018\textwidth}p{0.018\textwidth}|p{0.018\textwidth}p{0.018\textwidth}p{0.018\textwidth}?}
\Xhline{3\arrayrulewidth}
\textbf{Threat} &
\textbf{Impact} &
\multicolumn{5}{c|}{\textbf{Threat Model}} &
\multicolumn{3}{c?}{\makecell{\textbf{Vulnerable} \\ \textbf{Anti-Malware} \\ \textbf{Solutions}}}\\
\cline{3-10}
& & 
    \rotatebox{90}{Scope}&
    \rotatebox{90}{Capability} &
    \rotatebox{90}{MITM} &
    \rotatebox{90}{OPSAM} &
    \rotatebox{90}{OP} & 

   \rotatebox{90}{McAfee’s GTI} &
   \rotatebox{90}{Tenable’s Nessus} &
   \rotatebox{90}{Team Cymru's MHR} \\
   
\Xhline{3\arrayrulewidth}

\multirow{2}{*}{Silencing} & \multirow{2}{*}{Malware running on the target machine} & \CC{30} \textbf{SP} & \CC{30} \textbf{LT} & \CC{30} \cmark & \CC{30} \cmark & \CC{30} \cmark & \CC{30} \cmark & \CC{30} \cmark & \CC{30} \cmark \\

& & \textbf{LS} & \textbf{CT} & \cmark & \xmark & \xmark & \cmark & \cmark & \cmark \\
\hline

\multirow{2}{*}{False alerts} & \multirow{2}{*}{Alerting legitimate file as malicious (may lead to legitimate file deletion)} & \CC{30} \textbf{SP} & \CC{30} \textbf{LT} &\CC{30} \cmark & \CC{30} \cmark & \CC{30} \cmark & \CC{30} \xmark & \CC{30} \cmark & \CC{30} \cmark  \\

& & \textbf{LS} & \textbf{CT} &  \cmark & \xmark & \xmark & \xmark & \cmark & \cmark  \\
\hline

Information disclosure  & Exposing files that exist within the target machine & \CC{30} \textbf{LS} &\CC{30} \textbf{CE} & \CC{30} \cmark &\CC{30} \xmark & \CC{30} \xmark &\CC{30} \cmark & \CC{30} \cmark &\CC{30} \cmark\\

\Xhline{3\arrayrulewidth}

\multicolumn{10}{l}{Attack scope -  \textbf{SP}: specific , \textbf{LS}: large scale.} \\

\multicolumn{10}{l}{Attacker's required capabilities -  \textbf{LT}: limited tampering, \textbf{CT}: complete tampering, \textbf{CE}: complete eavesdropping.}\\

\multicolumn{10}{l}{Attacker model -  \textbf{MITM}: man-in-the-middle, \textbf{OPSAM}: off-path with IP spoofing capability and an adjacent machine, \textbf{OP}: off-path.} \\
\end{tabular}

\caption{A summary of threats applicable to consumers of anti-malware solutions that make use of DNSAML services.}\label{tab:threats}

\end{table*}

Based on a previous threat analysis of the DNS~\cite{atkins2004threat} and recently published DNS cache poisoning attacks \cite{herzberg2012security,man2020dns,klein2020cross}, we consider three type of attackers that can acquire the abovementioned capabilities: a man-in-the-middle attacker, an off-path attacker with IP spoofing capability that controls an adjacent machine, and an off-path attacker without any additional capabilities.

Today, with the wide adoption of SSL for encrypting and signing web traffic, the capabilities of the different attackers are very limited.
A man-in-the-middle attacker may be able to redirect web traffic by spoofing DNS responses.
However, with SSL encryption in place, the attacker must perform a social engineering attack or exploit a vulnerability within the browser in order for the attack to be successful and go undetected.
In our case, the plain-text DNS protocol allows an attacker to carry out the attacks described below \emph{secretly} and without the need to exploit additional vulnerabilities.
This makes the above-mentioned capabilities extremely important and valuable to attackers.

\subsubsection{\textbf{Man-in-the-middle attacker (MITM)}}
In this attacker \\
model, we assume an attacker that controls a machine that resides on the path between the DNS resolver and DNS forwarders.
Since DNS queries and responses are transmitted in a single unsigned and unencrypted UDP packet, such an attacker can acquire complete DNS eavesdropping and tampering capabilities. 
Potential threat actors are adversaries that control the Internet backbone (such as autonomous systems and/or Internet service providers) and adversaries that have access to DNS resolvers or authoritative name servers operated by global DNS services. 
In addition, an attacker that controls a machine within the same network segment as the target machine can acquire a MITM capability by exploiting a vulnerability in the network stack (such as ARP poisoning).

\subsubsection{\textbf{Off-path attacker with IP spoofing capability that controls an adjacent machine (OPSAM)}}
In this attacker model, we consider the classic DNS cache poisoning attacker model~\cite{herzberg2012security,man2020dns}.  
Specifically, we assume an attacker with an IP spoofing capability that controls a machine that resides off the path between the DNS resolver queried by the agent and the DNSAML service.
Assuming an IP spoofing capability is plausible, since 30.5\% of autonomous systems (ASes) do not block spoofed IP packets~\cite{luckie2019network}, and an attacker must only find one node that can spoof IPs.
In addition, we assume the attacker controls an additional machine that resides within the local area network (LAN) of the resolver.
As demonstrated in a recent study, this attacker can implement a cache poisoning attack on the most popular DNS providers (such as Google, Cloudflare, OpenDNS, Comodo, Dyn, Quad9, and AdGuard)~\cite{man2020dns}.
This attack leverages a universal side-channel attack in the network stack to overcome source port randomization (which is the widely adopted defense against DNS cache poisoning).
Since the success of this attack depends on a race condition of the DNS response, it succeeds with some constant probability.
Accordingly, such an attacker can only acquire a limited DNS tampering capability. 

\subsubsection{\textbf{Off-path attacker (OP)}}
In this attacker model, we assume an off-path attacker without any additional capabilities. 
As demonstrated in a recent study, this attacker can implement a cache poisoning attack on Linux-based DNS revolvers~\cite{klein2020cross}).
This attack exploits a vulnerability within the pseudorandom number generator (PRNG) of the Linux operating system (as well as Android) to overcome source port randomization.
Since the success of this attack depends on a race condition of the DNS response, it succeeds with some constant probability.
Accordingly, such an attacker can only acquire a limited DNS tampering capability.

\subsection{\label{subsec:primary_threats}Primary Threats}
The use of DNSAML services by anti-malware solutions introduces three primary threats to the consumers of these solutions (summarized in Table~\ref{tab:threats}):

\subsubsection{\textbf{Silencing}}
The responses made by DNSAML services to agents indicate whether a scanned file is malicious. 
Accordingly, an attacker with limited or complete DNS tampering capability can spoof the DNSAML response cause an agent to incorrectly classify a malicious file. 
Such an attack will result in allowing the execution of malicious files on the victim machine.

This threat is significant because it allows an attacker to perform host-based manipulation without a host-based presence.
The extent of the threat is affected based on whether the attacker has a limited or complete DNS tampering capability though.
With a limited DNS tampering capability, an attacker will be able to allow the execution of specific malicious files (e.g., silencing an alert on a specific malware).
In comparison, with a complete DNS tampering capability, an attacker will be able to allow execution of any malicious file on a large scale.

Furthermore, this threat is significant because it goes undetected.
For example, a MITM attacker can entirely disable all queries to a DNSAML service to prevent an agent from scanning files in general.
However, the complete disablement of DNS queries to a DNSAML services is expected to result in displaying an error.
Conversely, by spoofing the DNS response of the DNSAML service, such an attack succeeds while going undetected.

\subsubsection{\textbf{False alerts}}
The responses made by DNSAML services to agents indicate whether a scanned file is malicious. 
Accordingly, an attacker with limited or complete DNS tampering capability can spoof the DNSAML response cause an agent to incorrectly classify a benign file as malicious. 
Such an attack will result in alert fatigue, and potential sanctions against a benign file.
Furthermore, the inaccurate reports resulted by this attack, may damage the brand of the anti-malware solution, and the trust of its consumers. 

Similarly to the silencing threat, the significance of the false alert threats stems from the ability to carry it on a host-machine without host presence, and go undetected.

\subsubsection{\textbf{Sensitive information disclosure}}
The queries made by anti-malware agents to DNSAML services to agents provide information regarding scanned files.
Accordingly, an attacker with a DNS eavesdropping capability can intercept the queries made an agent to its DNSAML service to gain information about scanned files.
With this information, a threat actor spreading malware can (a) determine whether a specific malware was successfully delivered to a victim endpoint without being detected; (b) learn which known malware go undetected by a DNSAML service and deliver them to the victim; and (c) assess the security posture of an organization based on the number of file scans, their file information and their responses that indicate the rate of malicious downloaded files by that organization.

\section{\label{sec:in-depth} Attacks On DNSAML Services}
In this section, we describe the attacks that materialize the threats defined in Section~\ref{sec:threat_analysis}.
Then, we present a methodology for validating whether the attacks are applicable to three anti-malware agents via their use of a DNSAML service, namely McAfee's Global Threat Intelligence (GTI), Tenable's MalwareDB and Team Cymru’s Malware Hash Registry (MHR).
\new{The validation methodology consists of several challenges that limit our ability to scale it for additional solutions, as discussed in Section~\ref{sec:discussion}.}
Finally, we present the results of the attack validation in Subsection~\ref{subsec:Results}.

To provide better intuition regarding the results, in Subsection~\ref{subsec:in-depth}, we provide further details about the inner workings of the communication between the anti-malware agents and their DNSAML services to increase understanding as to why the attacks succeed or fail with specific agents.

\subsection{\label{subsec:attacks} From Threats to Attacks}
We actualize the threats defined in Section~\ref{sec:threat_analysis}, performing concrete attacks that validate the feasibility of the threats, and provide basic notations. \\

\noindent \textbf{Information Disclosure Attack (ATT-ID):} This attack actualizes the information disclosure threat described in Section~\ref{sec:threat_analysis}.
In this attack, an attacker with DNS eavesdropping capability seeks to learn whether \emph{specific} malware was scanned by the anti-malware agent in the victim machine.

The attack is constructed as follows:
\begin{enumerate}[leftmargin=*]
    \item The attacker constructs dictionary mapping files to their corresponding signatures (generated by the anti-malware agent): 
        \begin{enumerate}[leftmargin=*]
            \item The attacker learns which anti-malware agent is installed on the victim machine by intercepting the victim's outgoing DNS queries and matching the DNS zones to the DNSAML services described in Section~\ref{sec:usingdns}. 
            \item The attacker installs an anti-malware agent on a machine, similar to that installed on the victim machine. 
            \item The attacker deploys a set of malware on its machine and inspects the outgoing DNS traffic from the agent to the DNSAML service.
            \item The attacker constructs dictionary mapping files to signatures that appear in the outgoing DNS traffic from the agent to the DNSAML service.
        \end{enumerate}
        
    \item The anti-malware agent scans a malicious file.
    
    \item The attacker intercepts the victim machine's outgoing DNS traffic and matches every DNS query made to the DNSAML service with the set of signatures in the dictionary.
    \item Every successful match indicates that a DNS query was made, most likely by the victim machine's installed agent, indicating that the victim's machine contains malware from the set.
\end{enumerate}

\begin{figure}[t]
    \centering
    
    
    \subfigure[The false alert attack (as implemented on Tenable’s MalwareDB). \label{fig:false-alert-attack}]{\includegraphics[width=0.5\textwidth]{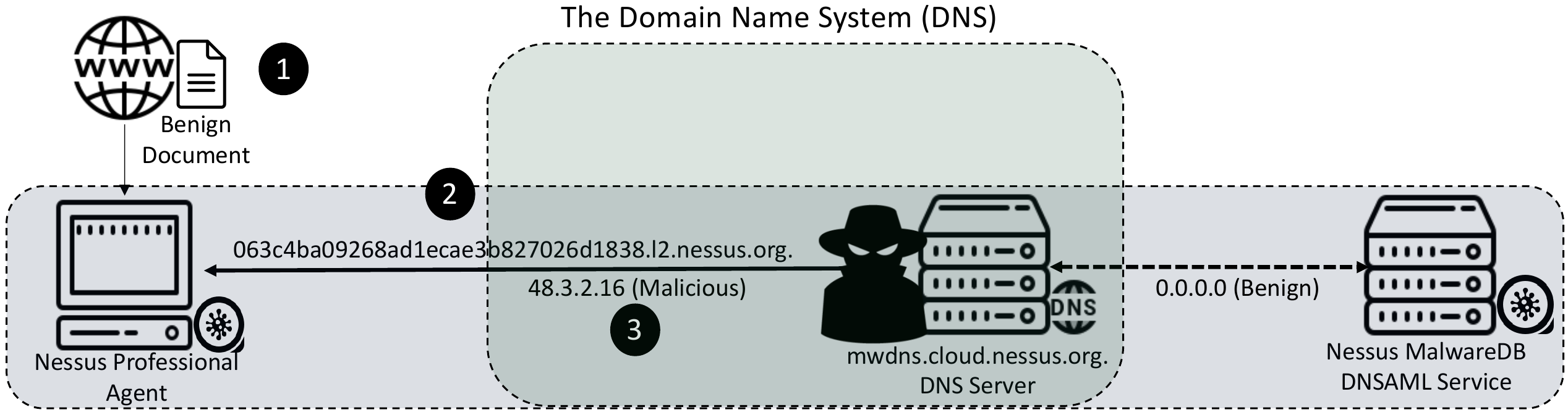}}\\
    
    \subfigure[The silencing attack (as implemented on McAfee's GTI). \label{fig:disable-protection}]{\includegraphics[width=0.5\textwidth]{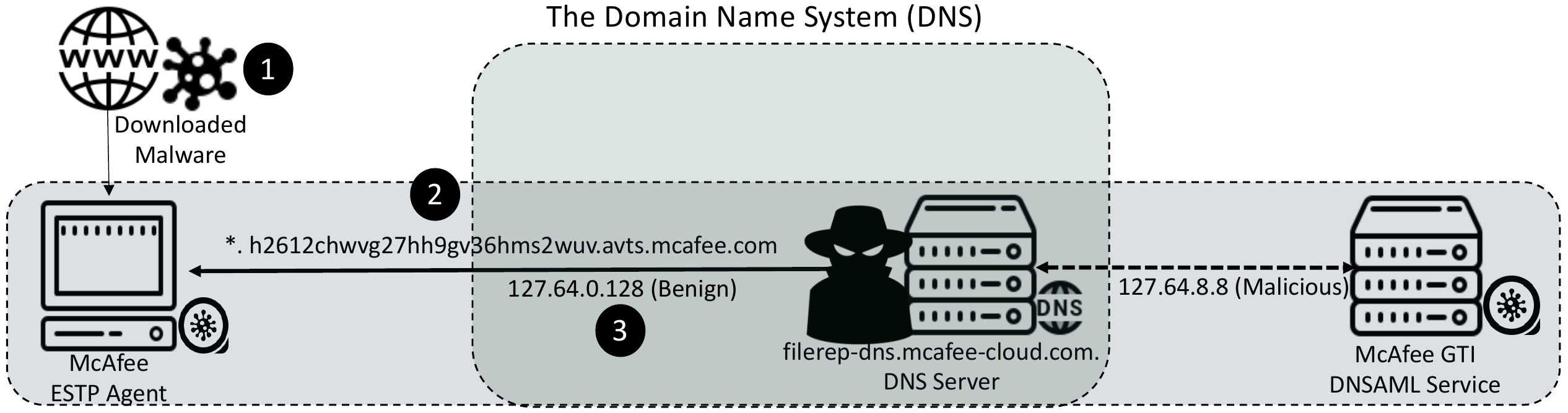}}
    
\caption{An illustration of attacks on DNSAML services: the false alert attack (a) and the silencing attack (b).
In both attacks, an anti-malware agent downloads a file (1) and queries its DNSAML service to learn whether the file is malicious (2). The attacker spoofs the DNSAML service response to convince the agent that a malicious file is incorrectly benign or vice versa (3).}
\label{fig:attack_scheme}

\end{figure}

\noindent \textbf{False Alert Attack (ATT-FA):}
This attack actualizes the information disclosure threat described in Section~\ref{sec:threat_analysis}.
In this attack, an attacker with DNS tampering capability (full or limited) causes the anti-malware agent installed on a victim machine to classify a benign file as malicious.

The attack is illustrated in Figure~\ref{fig:false-alert-attack} as follows:
\begin{enumerate}[leftmargin=*]
    \item The anti-malware agent installed on the victim's machine scans a benign file (for instance, a file downloaded from the Internet as shown in step 1).
    \item The anti-malware agent scan results in a DNS query to the DNSAML service with the benign file signature (step 2).
    \item The attacker spoofs the DNSAML service's response to a \emph{malicious response} i.e., a DNS response to which the anti-malware agent is designed to act with a \emph{malicious action} such as: alert, delete, or quarantine (step 3).
    \item The anti-malware agent incorrectly performs the malicious action on the file.
\end{enumerate}

\noindent \textbf{Silencing Attack (ATT-S):}
In this attack, an attacker with DNS tampering capability (full or limited) causes the anti-malware agent installed on a victim machine to classify a malicious file as benign.

The attack is illustrated in Figure~\ref{fig:disable-protection} as follows:
\begin{enumerate}[leftmargin=*]
    \item The anti-malware agent installed on the victim's machine scans a malicious file (for instance, a malicious file downloaded from the Internet as shown in step 1).
    \item The anti-malware agent scan results in a DNS query to the DNSAML service with the malicious file signature (step 2).
    \item The attacker spoofs the DNSAML service's response to the \emph{benign response}. i.e., a DNS response to which the anti-malware agent is designed to act with a \emph{benign action} such as: ignore (step 3).
    \item The anti-malware agent incorrectly performs the benign action on the file.
\end{enumerate}

\subsection{\label{subsec:validation} Experimental Methodology}
In this section, we describe the experimental methodology used to validate the successful implementation of the above-mentioned attacks on specific anti-malware agents.

We begin by defining the following notations:
\begin{itemize}\leftmargin=1em
    \item $F_{B}$: a set of files that are considered by the anti-malware agent to be benign.
    \item $F_{M}$:  a set of files that are considered by the anti-malware agent to be malicious.
    \item $R_{O}$: the set of allowed responses by the DNSAML service
    \item $A_{B}$: the set of actions taken by the anti-malware agent for benign files.
    \item $A_{M}$: the set of actions taken by the anti-malware agent for malicious files.
\end{itemize}

For clarification, any file can \new{be} assigned into either $F_{B}$ or alternatively $F_{M}$, based on whether the anti-malware solution scan results in an action from $A_{B}$ or $A_{M}$ correspondingly.

\noindent\textbf{Validating the information disclosure attack (\textit{\textbf{ATT-ID}}):}
We installed the anti-malware agent on two endpoints.
Then, we deploy a file on both endpoints and observe the outgoing DNS queries to the DNSAML service.
We assert that the information disclosure attack is feasible for a specific DNSAML service if and only if the file signatures within the outgoing DNS queries made from both endpoints match exactly, i.e., the signature are independent of the endpoint and depend only on the scanned file.
This assertion implies that a DNS eavesdropping attacker will be able to successfully match a set of malware signatures in intercepted DNS traffic, because the file signatures on the and victim setup are identical.

The validation process of the information disclosure attack is described below:

\begin{tcolorbox}[left=0pt,breakable,title=\textit{Information Disclosure Attack Validation}]
\small
    \begin{enumerate}\leftmargin=1em
        \item Let $e1, e2$ be two different endpoints with the same anti-malware agent installed.
        \item For every file $f^{i}$ in $F_{B} \cup F_{M}$:
        \begin{enumerate}
            \item Deploy the $f^{i}$ on $e1$ and record the first DNS query made to the DNSAML service: $q(f^{i}, e1)$
            \item Deploy the $f^{i}$ on $e2$ and record the first DNS query made to the DNSAML service: $q(f^{i}, e2)$
            \item If the file signatures on $q(f^{i}, e2)$ and  $q(f^{i}, e2)$ mismatch, return False
        \end{enumerate}
        \item Return True
    \end{enumerate}
\end{tcolorbox}

\noindent\textbf{Validating the false alert attack (\textit{\textbf{ATT-FA}}):}
We deploy benign files on an endpoint protected by an anti-malware agent associated with the inspected DNSAML service.
For every deployed file, we record the DNS response returned from the DNSAML service and verify that the agent's action regarding the file is a benign action (i.e., a sanity check).
Following that, we redeploy each file when spoofing the DNS response returned by the DNSAML service and check whether the action changed from a benign action to a malicious one.
Specifically, the set of spoofed responses is extracted from the set of responses of the DNSAML service on our dataset, therefore limiting the search.
We claim that an attacker capable of DNS tampering can launch the false alert attack if and only if for every file at least one spoofed response results in changing the agent's action from a benign action to a malicious one.

The validation process of the false alert attack is described below:

\begin{tcolorbox}[left=0pt,title=\textit{False Alert Attack Validation}]
\small
    \begin{enumerate}\leftmargin=1em
        \item Let $V=\emptyset$ be the set of benign files for which the agent raised a false alert
        \item For every file $f^{i}_{B}$ in $F_{B}$:
        \begin{enumerate}
            \item Deploy the $f^{i}_{B}$ on the endpoint
            \item Record the DNSAML service response $r(f^{i}_{B})$
            \item Record the agent's action regarding the file: $a(f^{i}_{B}, r(f^{i}_{B}))$
            \item Verify $a(f^{i}_{B}, r(f^{i}_{B})) \in A_{B}$; otherwise, return False
            \item For every response $r_{j}$ in $R_{O}$:
            \begin{enumerate}
                \item Deploy the $f^{i}_{B}$ on the endpoint
                \item Spoof the DNS response to $r_{j}$ instead of $r(f^{i}_{B})$
                \item Record the action : $a(f^{i}_{B}, r_{j})$
                \item If $a(f^{i}_{B}, r_{j}) \in A_{M}$, then $V = V \cup \{f^{i}_{B}\}$
            \end{enumerate}
        \end{enumerate}
        \item Return True if $V = F_{B}$
    \end{enumerate}
\end{tcolorbox}

\noindent\textbf{Validating the silencing attack (\textit{\textbf{ATT-S}}):} We deploy malicious files on an endpoint protected by the anti-malware agent associated with the inspected DNSAML service.
For every deployed file, we record the DNS response returned from the DNSAML service and verify that agent's action regarding the file is a response to a malicious file (i.e., a sanity check).
Following that, we redeploy each file when spoofing the DNS response returned by the DNSAML service and check whether the action changed from a malicious action to a benign one.
Specifically, the set of spoofed responses is extracted from the set of responses of the DNSAML on our dataset, therefore limiting the search.
We claim that an attacker capable of DNS tampering can launch the silencing attack if and only if for every file at least one spoofed response results in changing the agent action from a malicious action to a benign one.

The validation process of the \new{silencing} attack is described below:

\begin{tcolorbox}[left=0pt,breakable,title=\textit{Silencing Attack Validation}]
    \small
    \begin{enumerate}\leftmargin=1em
        \item Let $V=\emptyset$ be the set of malicious files which the attacker caused the agent to ignore
        \item For every file $f^{i}_{M}$ in $F_{M}$:
        \begin{enumerate}
            \item Deploy the $f^{i}_{M}$ on the endpoint
            \item Record the DNSAML service response $r(f^{i}_{M})$
            \item Record the agent's action regarding the file: $a(f^{i}_{M}, r(f^{i}_{M}))$
            \item Verify $a(f^{i}_{M}, r(f^{i}_{M})) \in A_{M}$; otherwise, return False
            \item For every response $r_{j}$ in $R_{O}$:
            \begin{enumerate}
                \item Deploy the $f^{i}_{M}$ on the endpoint
                \item Spoof the DNS response $r_{j}$ instead of $r(f^{i}_{M})$ 
                \item Record the action : $a(f^{i}_{M}, r_{j})$
                \item If $a(f^{i}_{M}, r_{j}) \in A_{B}$, then $V = V \cup \{f^{i}_{M}\}$
            \end{enumerate}
        \end{enumerate}
        \item Return True if $V = F_{M}$
    \end{enumerate}
\end{tcolorbox}

\subsection{\label{subsec:experimental_setup} Experimental Setup}
The experimental setup includes a Victim machine, an Attacker machine, and a dataset of benign and malicious Files.

\noindent \textbf{Victim Machine:} A virtual machine (VM) with the recent Windows 10 operating system.
The VM is connected to the Internet to allow the installed agent to query its DNSAML service.
In each experiment, we install the most recent version of the inspected malware agent, namely:
(1) McAfee Endpoint Security Threat Prevention (ESTP) version 10.6.1.1386.8, (2) Nessus Professional Version 8 (8.11.1), or, (3) python-tools-cymru~\cite{python-cymru-tools}.

\noindent \textbf{Attacker Machine:} A virtual machine running Kali Linux. 
The attacker machine resides as a MITM between the victim machine and the Internet. 
The DNS interception and spoofing were conducted with the Ettercap tool (for both ARP poisoning and DNS spoofing).

\noindent \textbf{Dataset of Benign and Malicious Files:}
The set of benign files includes 10 new Microsoft Office documents of various formats that we created and could confirm to be benign.
To construct the set of malicious files, we had to obtain files that the inspected anti-malware solution ``believes'' to be malicious.
This requirement was critical to guarantee that the DNSAML service's response instructs the agent to perform the malicious action. 

\begin{figure}[t]
    \centering
    \includegraphics[width=0.45\textwidth]{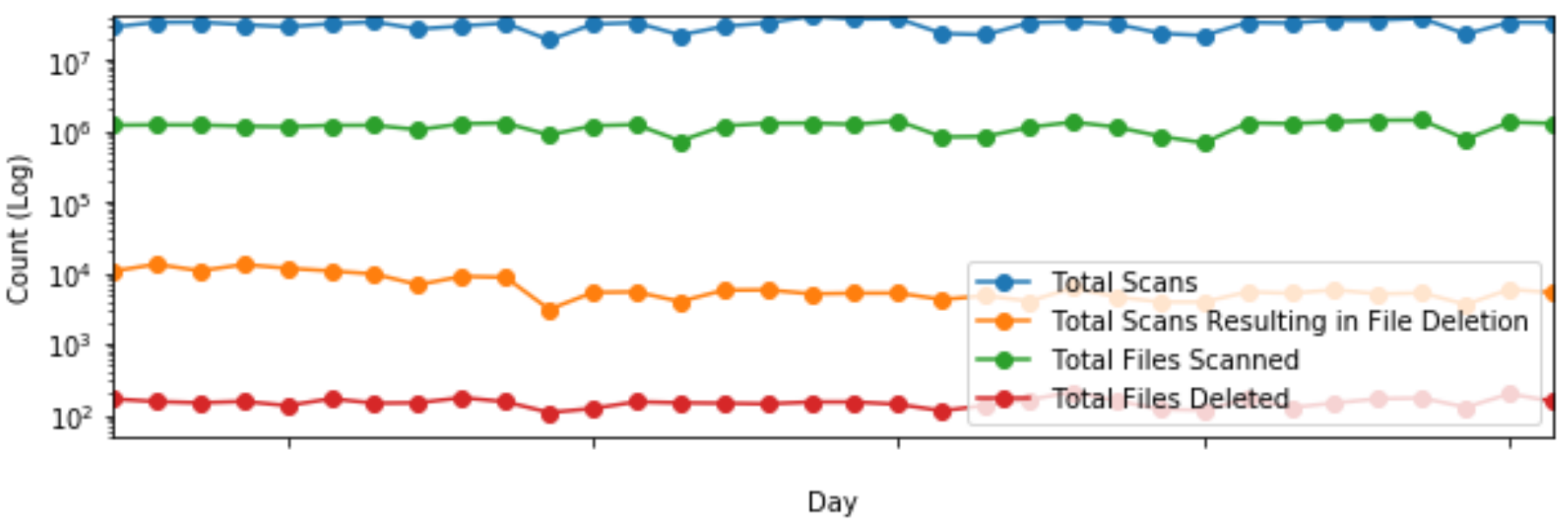}
    \caption{The number of total and unique file scans by McAfee GTI and those resulting in a deletion response.}
    \label{fig:mcafee_deletions}
\end{figure}

In the case of Tenable's MalwareDB, we extracted MD5 file hash signatures that we observed within the DNS traffic whose DNS response was extremely rare (i.e., less than 0.1\% of all DNS queries).
For instance, the query \url{b48eb0932dfb629ce70d7142ceb74a13.l2.nessus.org.} was observed only once within the 30-day period, and therefore we extracted the file hash signature ``b48eb0932dfb629ce70d7142ceb74a13.''
In turn, we used the VirusTotal service to download the malicious files based on its MD5 file hash.
Overall, we used seven files that we were confident the Nessus agent would classify as malicious. \\
In the case of McAfee GTI, we faced greater difficulty, since the file hash algorithm used was not MD5 or SHA1, which are the known standard algorithms for computing malicious file hashes.
Accordingly, we were unable to download malicious files directly from VirusTotal.
Instead, we carefully modified two malicious binaries to form 10 files which we manually confirmed to be new and were classified as malicious by the DNSAML service. \\
In the case of Team Cymru, we used the example of malicious file hash signatures demonstrated in their guidelines~\cite{cymru_mhr}.

\subsection{\label{subsec:Results} Results}
The three attacks, namely ATT-ID, ATT-FA, and ATT-S, were evaluated on the three anti-malware solutions.
The results are presented in Table~\ref{tab:threats}.
We also provided two screenshots and four videos that demonstrates the attack validations in Appendix~\ref{sec:screenshots}. 
The first screenshot (see Figure~\ref{fig:nessus-fa-screenshot}) shows the DNS responses spoofed by the attacker, and the Nessus professional scan report that incorrectly lists all of the system legitimate files as malicious, thus resulting creating a case of alert fatigue for the victim.
The second screenshot (see Figure~\ref{fig:mcafee-silencing-screenshot}) shows the differences between the scanning of a malicious file without attack and with the silencing attack on the McAfee ESTP agent.
Without the attack, the Artemis Trojan is successfully detected by the agent, that takes the action of deleting the Trojan file.
In contrast, when the silencing attack is carried, the attacker spoofs the DNS responses made by McAfee's GTI DNSAML service to the agent, which results in an empty scan report. 
As a result, the malicious Artemis Trojan, is ignored by the agent and is allowed to be executed on the victim machine.
The videos are available on the following private and anonymous YouTube link \footnote{\url{https://youtube.com/playlist?list=PLMhL8ch_vmMBRju0mSA_997WgBp7TK5KU}}.

Our primary findings when validating the attacks on the experimental setup were that the inspected anti-malware solutions that make use of DNSAML services lack additional safeguards to compensate for the insecurities of the DNS protocol.
In particular, the lacking safeguards are as follows:
\begin{enumerate}[leftmargin=*]
    \item\textbf{Lack of encryption.} The DNS queries made by all anti-malware agents and and their DNSAML services weren't encrypted. 
    \item\textbf{Lack of authentication.} The DNS responses anticipated by all anti-malware agents were not authenticated to ensure the origin is the DNSAML service.
    \item\textbf{Static file signature.} The file signature in all the DNS queries was unchanged between sessions and different scanning machines.
    In some solutions, (e.g., Tenable's Nessus) the signature publicly reveals the scanned file.
    \item \textbf{Errors are not propagated.} In some solutions (e.g., Tenable's Nessus), connectivity errors and encoding errors are not propagated to the operator, thus limiting the service provider ability to infer it is attacked.
\end{enumerate}

\subsection{\label{subsec:in-depth}In-Depth Analysis}
In this subsection, we provide further details about the inner workings of the communication between the anti-malware agents and their DNSAML services to increase understanding as to why the attacks succeed or fail with specific agents.

\noindent\textbf{McAfee’s GTI:}
McAfee's ESTP agent queries the GTI DNSAML service when it identifies a suspicious file that does not trigger existing signature DAT files~\cite{mcafee_gti}.
The DNS query is issued to one of two possible DNS zones owned by McAfee, namely \url{avts.mcafee.com} or \url{avqs.mcafee.com}, as discussed in Section~\ref{sec:usingdns}.
Each query specifies a 32-bit signature of the scanned file and requests an IPv4 address response that encodes the predetermined action for the scanned file.
In our experiments, the 32-bit signature never matched the VirusTotal service.
Therefore, we suspect that the signature is the product of an internal specification.


The DNS response is an IPv4 address.
Using the dataset described in Subsection~\ref{subsec:dataset}, we were able to identify 25 unique IPv4 responses.
Only one of these IPv4 responses results in the deletion of a scanned file by the agent, presumably because it was classified as malicious --- 127.64.8.8.
Therefore, we conclude that the set of malicious responses includes only this IPv4, which we later refer to as \emph{the deletion response}. 
When the deletion response is returned, the ESTP agent engages in an additional DNS query.
The additional DNS query requests a DNS TXT record, typically used to store long textual responses in the domain name system.
The content of the response is encrypted and encoded using a Base64 encoding scheme.
Within the scope of this research, we were unable to determine exactly what is encoded within the response, however we found that if the response is altered or blocked, the scanned file is not deleted despite being malicious.

With the knowledge of the deletion response, we were able to accurately determine the rate of files deleted by McAfee's endpoint solutions based on the rate of the deletion response compared to the rest of the responses, as shown in Figure~\ref{fig:mcafee_deletions}.

\noindent\textbf{Tenable’s MalwareDB:}
The Nessus Professional agent has a built-in malware scanning feature that is triggered either manually or periodically. 
The agent acts only as a scanner that generates a scan report, without the capability of applying sanctions on malicious files.
While scanning, Nessus issues DNS queries to the MalwareDB server for a malware hash lookup of suspicious files.

When the malware scan starts, the agent sends a DNS query to \url{chk.l2.nessus.org} to confirm connectivity to its servers, before proceeding with any malware lookups. 
Once connectivity has been confirmed, the scanner issues a DNS query for every scanned file by prepending a custom 32-bit hash of the scanned file (i.e., \emph{Nessus hash}) to the \url{l2.nessus.org} domain.
The DNS query requests an IPv4 address resource record and in response, receives one that encodes the results of the malware hash lookup.

Every malware scan results in a scan report.
Within the scan report, every file that is identified as malicious is associated with the number of security engines by which it was scanned and the number of engines that classified the file as malicious. In some cases, the report also indicates which particular engines considered the file as malicious.
In addition, the scan report assigns each malicious file with a report that users can browse by visiting a URL.
The URL is \url{http://malwaredb.nessus.org/malware/} followed by the file hash.
The report contains the MD5 hash of the scanned file, as well as a detailed description of the malware, thus allowing an eavesdropping attacker to obtain a standard hash that can be matched against other services.

In our attack validation, we observed that benign file scans always responded by Tenable's MalwareDB with one of two possible DNS responses.
Therefore, we refer to these responses as \emph{benign responses}.
Conversely, the rest of the IP addresses responded were shown to be malicious responses, as demonstrated in the attack validation results and scan report.

Furthermore, we found consistencies between labels appearing in the malicious responses and attributes appearing in the scan report.
Based on that, we were able to reconstruct a function that associates the report results to the IP address response, and vice versa, thus allowing an eavesdropping attacker to obtain the report results simply by applying the function to the IP addresses contained in the responses.

The function that maps IP addresses contained in the responses to the report results is defined as follows: (a) The first three (MSB) bits (left-most) of the first octet  are ignored; (b) The remaining five bits of the first octet form a binary representation of the total number of engines that scanned the file; (c) The last five (LSB) bits of the second octet form a binary representation of the number of engines that classified the file as malicious; and (d) The third and fourth octet represent 16 ordered flags, each for a specific engine that classified the file as malicious.

\noindent\textbf{Team Cymru’s Malware Hash Registry:}
The inner workings of Team Cymru's MHR are fairly straightforward and well documented compared to the other solutions~\cite{cymru_mhr}.
In order to initiate a lookup for a potentially malicious file, the informal agent must first calculate the file's hash value using the MD5 or SHA1 algorithms.
Next, the agent issues a DNS query for either an IPv4 address or a text record, prepended with the file's hash value as the DNS resource record for the registered domain name: \url{malware.hash.cymru.com}, as if the hash value is a subdomain.
In the case of a benign file, a standard NXDOMAIN is returned as the \emph{benign response}.
In the case of a malicious file, the \emph{malicious response} is the IP address 127.0.0.2, as documented on~\cite{cymru_mhr}.

\section{\label{sec:countermeasures}Proposed Countermeasures} 
In this section, we review a set of countermeasures to prevent attacks stemming from the use of DNSAML services (see Section~\ref{sec:in-depth}).
We start by defining criteria to evaluate the countermeasures (subsection~\ref{subsec:countermeasures_methodology}); then we propose countermeasures and review them according to the criteria (subsection~\ref{subsec:countermeasures}).
A summary of the countermeasures is presented in subsection~\ref{subsec:countermeasures_summary}.

\begin{table*}[ht]
\centering
\resizebox{\textwidth}{!}{%
\begin{tabular}{|c|c|c|c|c|c|c|c|c|c|}
\hline
                  & \multicolumn{4}{c|}{Security}       & \multicolumn{2}{c|}{Performance}     & \multicolumn{2}{c|}{Compatibility} & Flexibility        \\ \hline
Countermeasure & \textbf{ATT-FA} & \textbf{ATT-FA} & \textbf{ATT-S} & \begin{tabular}[c]{@{}l@{}}Known \\ Limitations\end{tabular} & \begin{tabular}[c]{@{}l@{}}Response Time \\ Overhead \end{tabular} & Caching & \begin{tabular}[c]{@{}l@{}}Negligible \\ Infra. Changes\end{tabular}     & \begin{tabular}[c]{@{}l@{}}Negligible \\ Consumer Effort\end{tabular}    & \begin{tabular}[c]{@{}l@{}}Character \\ Limitation\end{tabular} \\ \hline
\begin{tabular}[c]{@{}l@{}}DNSAML services \\ without implementing \\ additional safeguards\end{tabular}                  & \Circle  & \Circle  & \Circle  &  & \Circle & \CIRCLE & \CIRCLE  & \CIRCLE       & \Circle  \\ \hline
Application-layer signing & \Circle & \CIRCLE & \CIRCLE &  & \LEFTcircle & \CIRCLE & \Circle  & \CIRCLE       & \Circle  \\ \hline
DoT~\cite{zhu2015connection} & \CIRCLE & \CIRCLE & \CIRCLE & ~\cite{houser2019investigation, bushart2020padding} & \LEFTcircle & \CIRCLE & \LEFTcircle & \LEFTcircle & \Circle  \\ \hline
DoH~\cite{rfc8484}                          & \CIRCLE & \CIRCLE & \CIRCLE & \cite{huang2020comprehensive} & \LEFTcircle & \Circle  & \LEFTcircle & \Circle        & \Circle  \\ \hline
REST APIs~\cite{wiki:restapi}                    & \CIRCLE & \CIRCLE & \CIRCLE &  & \Circle & \Circle  & \Circle & \Circle        & \CIRCLE \\ \hline
\Xhline{3\arrayrulewidth}
\multicolumn{10}{l}{Security -  \textbf{ATT-ID}: Information Disclosure attack , \textbf{ATT-FA}: False Alert attack, \textbf{ATT-S}: Silencing attack.} \\
\multicolumn{10}{l}{Fully satisfies (\CIRCLE), partially satisfies (\LEFTcircle), does not satisfy (\Circle).} \\
\end{tabular}%
}
\caption{Summary of proposed countermeasures to limit the consequences of attacks stemming from the insecure use of DNSAML services on anti-malware solutions and their consumers.}\label{tab:countermeasures}
\end{table*}

\subsection{\label{subsec:countermeasures_methodology} Assessment Criteria}
We define four categories of assessment criteria: security, performance, compatibility, and flexibility.
The following symbols (\CIRCLE,\LEFTcircle,\Circle) are used to indicate whether a criterion is fully satisfied, partially satisfied, or not satisfied by a countermeasure.

\subsubsection{\textbf{Security.}} The criteria in this category examine the level of protection provided by a countermeasure against the threats described in Section~\ref{sec:threat_analysis}.
Some of the countermeasures suffers from known security limitations (e.g., vulnerabilities under different attack models).
In these cases, we explicitly mention the security limitations for consideration when describing the countermeasures in subsection~\ref{subsec:countermeasures}, and in the summary table (see Table~\ref{tab:countermeasures}. 
\begin{list}{\textbf{Criterion \arabic{criteria_counter}.}}
{
\usecounter{criteria_counter}
\setlength\labelwidth{0.05in}
\setlength\leftmargin{0.11in}
}
\item \textit{Protection against the information disclosure attack:}\\
    (\CIRCLE) - The countermeasure provides end-to-end protection against the information disclosure attack.\\
    (\Circle) - The countermeasure does not provide end-to-end protection against the information disclosure attack.
    
\item \textit{Protection against the false alert attack:}\\
    (\CIRCLE) - The countermeasure provides end-to-end protection against the false alert attack.\\
    (\Circle) - The countermeasure does not provide end-to-end protection against the false alert attack.
    
\item \textit{Protection against the silencing attack:}\\
    (\CIRCLE) - The countermeasure provides end-to-end protection against the silencing attack.\\
    (\Circle) - The countermeasure does not provide end-to-end protection against the silencing attack.
    
\item \textit{Existence of any known security limitations:}\\
    (\CIRCLE) - We were unable to confirm the existence of any known security limitations of the countermeasure.\\
    (\Circle) - We confirmed the known security limitations of the countermeasure.
\end{list}

\subsubsection{\textbf{Performance.}} The criteria in this category examine the overhead incurred by using the proposed countermeasure. 
We find this category to be extremely critical, since DNSAML services are often valued for their low latency due which is due to a fast response time of the UDP transport layer and their ability to cache answers locally~\cite{cymru_mhr}.

\begin{list}{\textbf{Criterion \arabic{criteria_counter}.}}
{
\usecounter{criteria_counter}
\setlength\labelwidth{0.05in}
\setlength\leftmargin{0.11in}
\setcounter{criteria_counter}{4}
}

\item \textit{Response time overhead:}\\
    (\CIRCLE) - The countermeasure does not add overhead to the response time.\\
    (\LEFTcircle) - \new{The countermeasure adds marginal overhead to the response time as described in~\cite{bottger2019empirical} and/or can outperform plain-text DNS with minor improvements as described in~\cite{hounsel2020comparing}}.\\
    (\Circle) - The countermeasure adds a significant overhead to the response time.
    
\item \textit{Cache support:}\\
    (\CIRCLE) - The countermeasure has native support for caching.\\
    (\Circle) - The countermeasure does not have native support for caching.
\end{list}

\subsubsection{\label{subsubsec:compatibility} \textbf{Compatibility.}} The criteria in this category examine the difficulty of implementing the countermeasure and the effort required by consumers to integrate it. 
The motivation for including this criterion stems from the fact that anti-malware solutions are sometimes installed on servers with a strict network policy, such as email gateways or secure web gateway (SWG) proxies. 
In such scenarios, consumers that anticipate malware tend to reduce the attack surface on the server by allowing a limited set of Internet protocols, for instance, allowing outbound DNS but blocking outbound HTTP/S. Given the use of DNSAML services, we know that DNS is allowed by definition. 
Allowing encrypted DNS will therefore require some effort, and other protocols may be more difficult or infeasible under strict network policies.

\begin{list}{\textbf{Criterion \arabic{criteria_counter}.}}
{
\usecounter{criteria_counter}
\setlength\labelwidth{0.05in}
\setlength\leftmargin{0.11in}
\setcounter{criteria_counter}{6}
}
\item \textit{Infrastructure changes:}\\
    (\CIRCLE) - The anti-malware solution provider can implement the countermeasure without infrastructure changes.\\
    (\LEFTcircle) - The anti-malware solution provider can implement the countermeasure subject to negligible changes to the infrastructure (e.g., changes to the DNS resolver).\\
    (\Circle) - The anti-malware solution provider can implement the countermeasure subject to non-negligible changes to the infrastructure (e.g., exchange information using methods other than DNS, add safeguards to the application layer).
    
\item \textit{Consumer integration effort:}\\
      (\CIRCLE) - The consumer can integrate the countermeasure without any effort (e.g., upgrade the anti-malware agent).\\
    (\LEFTcircle) - The consumer can integrate the countermeasure subject to negligible effort (e.g., allow encrypted DNS traffic).\\
    (\Circle) - The consumer can integrate the countermeasure subject to non-negligible effort (e.g., allow outbound traffic to protocols other than the DNS protocol).
\end{list}

\subsubsection{\textbf{Flexibility.}} The criteria in this category examine the flexibility of sending additional information within an exchanged message.
Queried domain names (i.e., messages) over the DNS protocol are limited to 255 characters~\cite{mockapetris1987rfc1035}.
As a result, anti-malware solutions that rely on the DNS to report signatures cannot send additional information, unless they extend their service queries to a \emph{session} of queries.
Therefore, alternative Internet protocols that increase the character limit within a single message allow anti-malware solutions to send further information; for example, an anti-malware agent can send an entire binary to its service, thereby providing the richest form of information, as performed in the cases of cloud anti-virus solutions~\cite{oberheide2008cloudav, wiki:cloudav}.

\begin{list}{\textbf{Criterion \arabic{criteria_counter}.}}
{
\usecounter{criteria_counter}
\setlength\labelwidth{0.05in}
\setlength\leftmargin{0.11in}
\setcounter{criteria_counter}{8}
}

\item \textit{Character limitation:}\\
    (\CIRCLE) - The countermeasure supports at least 1 MB of exchanged information per scanned file.\\
   (\Circle) - The countermeasure supports less than 1 MB of exchanged information per scanned file.
\end{list}

\subsection{\label{subsec:countermeasures} Reviewed Countermeasures}
We review four countermeasures that are widely available and finalized from a technological readiness perspective, and are therefore suitable to overcome the insecurities of DNSAML effective immediately.
Conversely, we refrain from reviewing potential countermeasures that are either in draft stages during the writing of the study (e.g., DNS-over-QUIC~\cite{doq}) or listed an experimental (e.g., DNS over Datagram Transport Layer Security~\cite{rfc8094}).\\

\noindent\textbf{Application-layer signing} Application-layer signing allows remote services to sign responses returned over DNS with a private key.
In turn, every agent must verify the message's authenticity using a public key. 
From a security perspective, application-layer signing provides a robust countermeasure against data tampering attacks (i.e., false alert and silencing attacks).
However, it only partially protects against information disclosure attacks, because the data is still sent without encryption over the DNS protocol.
From a performance and flexibility perspective, the countermeasure uses the DNS protocol over the UDP transport layer, similarly to DNSAML services but with a additional payload (i.e., a signature).
The use of the DNS protocol also results in support for caching and a strict character limitation.
From a compatibility perspective, the countermeasure requires no changes by consumers.
However, the anti-malware countermeasure provider must design and integrate the additional safeguards of signing delivered messages and verifying the signatures of incoming messages.\\

\noindent\textbf{DNS over TLS (DoT)}~\cite{zhu2015connection} is privacy-preserving protocol based on a combination of the TCP stack and TLS that is used to improve DNS security and reliability. 
In DoT, DNS data is encrypted, which in turn reduces the impact of various DNS attacks (e.g., DNS hijacking) by establishing mutual connections. 
DNS based on TLS is designed to provide greater privacy, support large payloads, and mitigate hijacking and reflection distributed denial-of-service attacks more effectively than existing UDP protocols.
DoT provides a security countermeasure against all of the mentioned threats, by design.
The main security limitation of this countermeasure is privacy leakage~\cite{houser2019investigation, bushart2020padding}, allowing an MITM attacker to learn whether a user visits a specific set of domains, which is arguably a lesser concern than learning which \emph{exact} files are being scanned by a machine, but is still a concern.
\new{From a performance perspective, the use of the TCP transport layer results in reduced performance compared to the DNS protocol over the UDP transport layer, which is expected to become marginal, and even outperform plain-text DNS with the improvements proposed in~\cite{hounsel2020comparing}}.
However, caching is still supported, similarly to DNS without TLS. 
To support DoT in anti-malware solutions, consumers need to allow the DoT port (853), and solution providers must replace the DNS resolver with a DoT resolver.\\

\noindent\textbf{DNS-over-HTTPS (DOH)}~\cite{rfc8484} is a standard web protocol for sending DNS traffic over the privacy-preserving HTTPS. 
DoH provides a security countermeasure against all of the attacks mentioned.
Its main security limitation is its susceptibility to downgrade attacks~\cite{huang2020comprehensive} that force an client (e.g., agent) to fall back to plain-text DNS.
\new{From a performance perspective, the use of HTTP incurs a marginal overhead than to plain-text DNS~\cite{bottger2019empirical}, and caching is only supported at the recursive DNS server level and not within internal servers, as in the case of plain-text DNS.}
To support DoH in anti-malware solutions, consumers need to allow HTTPS traffic, and solution providers must replace the DNS resolver with a DoH resolver.\\

\noindent\textbf{REST APIs}~\cite{wiki:restapi} are a software architecture standard for interactive web applications.
Compared to the previously mentioned countermeasures, REST APIs replace the underlying DNS protocol to exchange file information with the HTTPS protocol.
The main upsides of REST APIs are security and flexibility, since compared to plain-text DNS, REST APIs allow larger messages (i.e., up to 2 MB with GET requests).
In contrast, REST APIs have the worst performance and compatibility issues of all of the reviewed countermeasures, because providers must change their infrastructure to support HTTPS, and consumers must permit HTTPS.

\subsection{\label{subsec:countermeasures_summary} Summary of Reviewed Countermeasures}
A summary of the reviewed countermeasures is presented in Table~\ref{tab:countermeasures}.
Each of the countermeasures provides varying levels of security, performance, compatibility, and flexibility.
Our interpretation of the trade-offs is as that (1) application-layer signing strongly favors performance and consumer compatibility over security, provider compatibility, and flexibility; (2) DoH and DoT provide a reasonable balance between performance and compatibility; and (3) REST APIs were found to be the least compatible of the countermeasures, but they provide ideal security and flexibility.
In their response to our responsible disclosure, McAfee indicated that it intends to enable REST APIs as their sole countermeasure, while Team Cymru plans on employing both plain-text DNS and REST APIs to assure backward compatibility.

\section{\label{sec:related_works} Related Work}
\noindent\textbf{Data exchange over the DNS protocol.}
The queries made to DNSAML services can be viewed as a special case of DNS tunneling~\cite{nadler2019detection,ahmed2019real,das2017detection,steadman2018dnsxd}, a method used to exchange information over the DNS protocol, typically to circumvent network policies.
This view is supported by Nadler et al.~\cite{nadler2019detection}, which mentions the use of DNSAML services as part of an analysis on DNS tunneling.
Therefore, this study on DNSAML services, can also be considered as work related to exploring \emph{legitimate} use cases of DNS tunneling, and threats associated with them --- a topic which arguably received insufficient attention.\\

\noindent\textbf{DNS privacy leakage.}
Several studies performed over the years have focused on DNS privacy issues and related solutions.
The majority of these studies addressed privacy-preserving protocols to improve the overall privacy of the DNS protocol and, as a result, limit leakage~\cite{castillo2009evaluation,lu2010towards,herrmann2014encdns,zhu2015connection}.
However, there are only a few studies (such as~\cite{krishnan2010dns}) that focused on privacy leakage of \emph{applications} as a result of using DNS, a topic which is critical due to the slow adoption of privacy-preserving protocols for DNS.
To the best of our knowledge, no study thus far has explored the privacy implications of legitimate applications (other than web browsers) that use the DNS protocol for data exchange or focused on anti-malware solutions, which are the subjects of this study.\\

\noindent\textbf{Practical attacks on the DNS protocol and its applications.}
Research on practical attacks on the DNS protocol and its applications has mainly examined general DNS protocol attacks, such as cache poisoning~\cite{dissanayake2018dns}, domain impersonation attacks~\cite{schwittmann2019domain}, amplification attacks~\cite{afek2020nxnsattack}, and DDoS attacks~\cite{alieyan2016overview,ahmed2017mitigating}.
However, studies regarding practical attacks on specific applications based on their use of the DNS protocol are rare.
In contrast, anti-malware solutions (e.g., antivirus), which are the focus of this study, are constantly evaluated for vulnerabilities due to their prevalence and the widespread trust placed in them ~\cite{quarta2018toward}.
To date, research has managed to identify various antivirus vulnerabilities~\cite{wressnegger2017automatically,al2011application,ormandy2011sophail}, however, exploiting antivirus vulnerabilities by manipulating DNS traffic, which is the topic of this study, has not been covered in prior research.

\section{\label{sec:discussion}Discussion}
\new{\textbf{Attack validations challenges and limitations:} The validation of the presented attacks for any anti-malware solution (see Subsec~\ref{subsec:validation}) consists of several challenges.}

\new{The first challenge is identifying all of the potential actions that the anti-malware solution may take for a scanned file, in order to categorize them into either the set of actions for malicious files ($A_M$) or for benign files ($A_B$).
To overcome this challenge, one must carefully read the anti-malware solution guide to configure its policy, as well as track undocumented behavior.
For instance: some of the anti-malware use DNSAML services to collect data for their internal analysis (e.g., ESET LiveGrid), rather than classify it according to a customer defined policy.}

\new{The second challenge deals with collecting a set of files that are classified as malicious ($F_M$) by the anti-malware solution's agent, without visibility to the anti-malware solution knowledge base i.e., a black-box setting.
To demonstrate that: our experiments included downloading several files that were considered as malicious by multiple security vendors on VirusTotal, and deploying them onto our setup to learn that they are either not classified as malicious by specific anti-malware solutions, or not resulting in an outgoing DNS query.}

\new{The third challenge involves to the difficulties in automating the process. 
Specifically, while it is easy to deploy a file and learn whether it was deleted, it is very difficult to learn if a file was quarantined or scanned without any threats because the anti-malware solutions do not provide an API.
All of these challenges combined reflect the high level of effort that is required to validate any anti-malware solution, thereby limiting the scalability of the validation process for all available anti-malware solutions.}

\section{\label{sec:conclusion}Conclusions and Future Work}
In this paper, we analyzed the security of DNS anti-malware list (DNSAML) services, to which anti-malware agents report suspicious files over the unsafe and privacy-lacking DNS protocol.
By examining a large-scale DNS log dataset, we identified 55 suspected DNSAML services used to deliver scanned file information.
Further analysis revealed that the use of DNSAML services is extremely prevalent, with more than 108 million daily scans performed by more than 2.85 million agents installed worldwide.

All of that points to the importance of performing a thorough security analysis of anti-malware solutions that make use of DNSAML services, in order to assess the threats introduced to anti-malware solution providers and their consumers.
To that end, we defined a threat model, under which attackers can perform three attacks: an information disclosure attack in which an attacker can learn which files are located on an endpoint machine, a false alert attack that results in incorrect security alerts for benign files, and a silencing attack that causes an an anti-malware agent to ignore known malicious files.
We demonstrated the feasibility of these attacks on three well-known anti-malware solutions. This confirmed our concerns regarding the exchange of file scan information over the insecure DNS protocol as the means of communication.

We also reviewed a set of four countermeasures that can assist anti-malware solution providers in limiting the consequences of attacks stemming from the insecure use of DNSAML services on their solutions and consumers.
The summary of countermeasures presented illustrates the clear trade-off between security, performance, compatibility, and flexibility that exists.
Our conclusion is that the DNS-over-HTTPS (DoH) and DNS-over-TLS (DoT) countermeasures are favorable from the compatibility and performance standpoints, while REST APIs are favorable from the security and flexibility standpoints.

In future work, we encourage research aimed at identifying additional DNSAML services, for instance using periodic queries over the DNS~\cite{daihes2020morton}, and the responsible disclosure of  anti-malware solution vulnerabilities as a result of the use as mentioned in this study.

\section{\label{sec:disclosure}Responsible Disclosure}


After establishing that the examined anti-malware solutions are vulnerable to the attacks described in this paper, we reached out to McAfee, Tenable, and Team Cymru, whose anti-malware solutions were analyzed in this research, via their official vulnerability disclosure email addresses and provided them with full disclosure of the suspected vulnerabilities, as well as the related technical details.\footnote{via security\_report@mcafee[.]com; vulnreport@tenable[.]com; and support@cymru[.]com; respectively.} 
\new{Following our disclosure, the providers acted as such:
\begin{enumerate}
    \item McAfee published a vulnerability report and an update fix 
    \item Team Cymru updated their documentation to encourage users to use their HTTPS services rather than the insecure DNS~\footnote{} as follows: ``Please be mindful of your risk tolerance and privacy concerns when choosing your transport protocol. DNS is convenient and a standard internet protocol, but does not normally afford the user integrity and confidentiality. HTTPS is recommended for those wanting increased integrity and confidentiality''.
    \item Tenable acknowledged the report and announced that they are not planning to make any immediate changes to the service at this time, but rather re-assess the situation for ways to improve in this scenario.
\end{enumerate} }

We note that some of the anti-malware solution providers chose to issue a response on the matter. 
More detailed information about our exchanges with the providers can be found in Appendix~\ref{sec:responsible_disclosure}.

\new{In addition to anti-malware solution providers that were investigated in this study (McAfee, Tenable, Team Cymru), we also reached out to other providers that were mentioned but not further investigated in this study, namely Sophos, Symantec and ESET.
We reached out more than two months prior to releasing the manuscript, to ensure these providers have sufficient time to look into our findings and consider changes to their services.}
\newpage

\bibliographystyle{ACM-Reference-Format}
\bibliography{sample}


\appendix

\appendixpage

\section{\label{sec:screenshots} Screenshots}
\begin{figure*}[!ht]
    \centering
    \includegraphics[width=\textwidth]{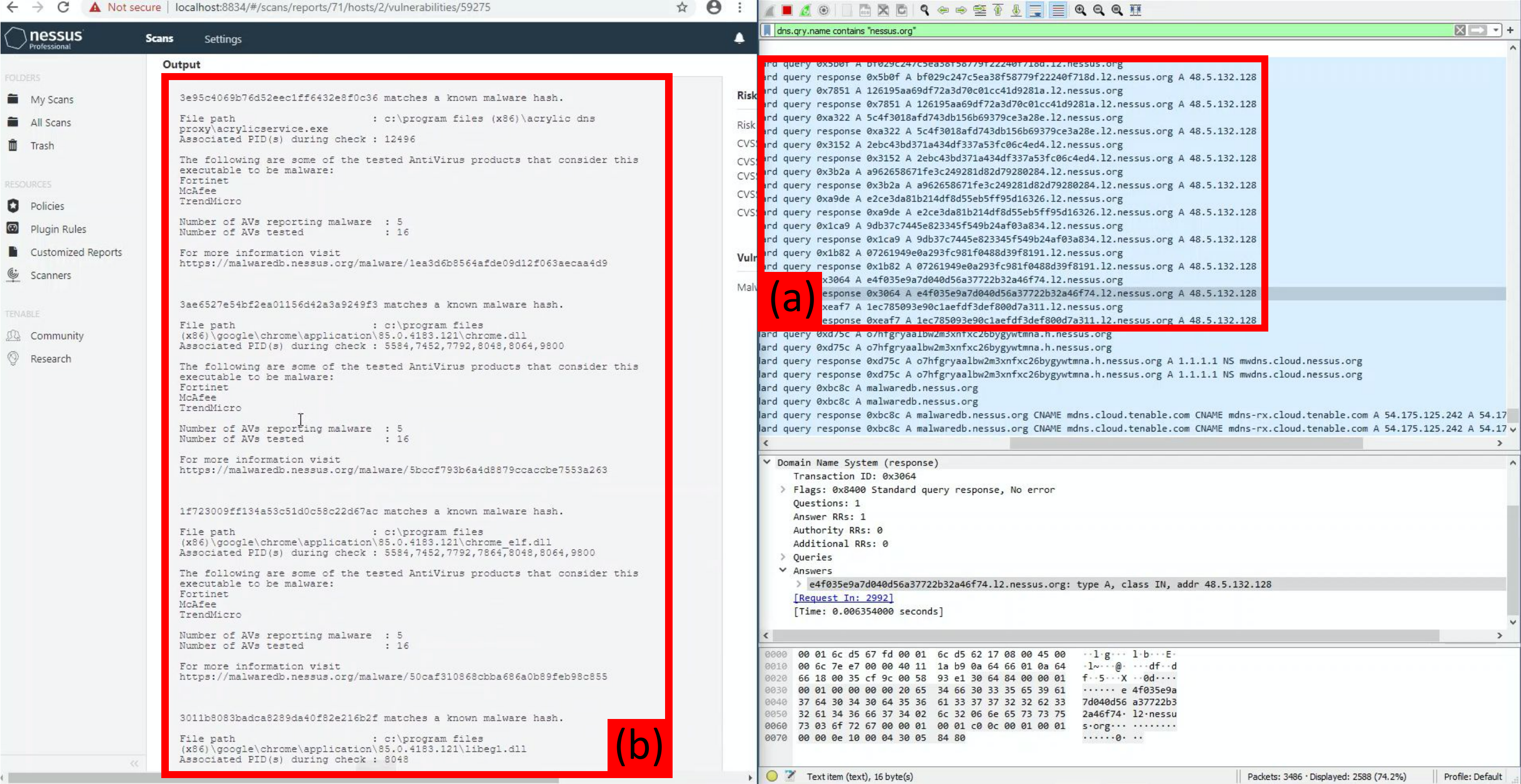}
    \caption{A screenshot the demonstrates the false alerts attack on the Tenable's Nessus Professional agent. 
    The victim initiates a full system scan using its installed agent.
    For every scanned file, the agent issues a DNS query to the MalwareDB DNSAML service. 
    A man-in-the-middle attacker spoofs the DNS responses on behalf of the MalwareDB DNSAML service to 48.5.132.128 (a).
    Based on the spoofed response, the agent incorrectly classifies all of the scanned files as malicious, and outputs false alerts (b).
    The outcome of the attack is that benign files (e.g., Chrome Web browser libraries) are listed as malicious, therefore creating an alert fatigue for the victim.}
    \label{fig:nessus-fa-screenshot}
\end{figure*}

\begin{figure*}[!ht]
    \centering
    \includegraphics[width=\textwidth]{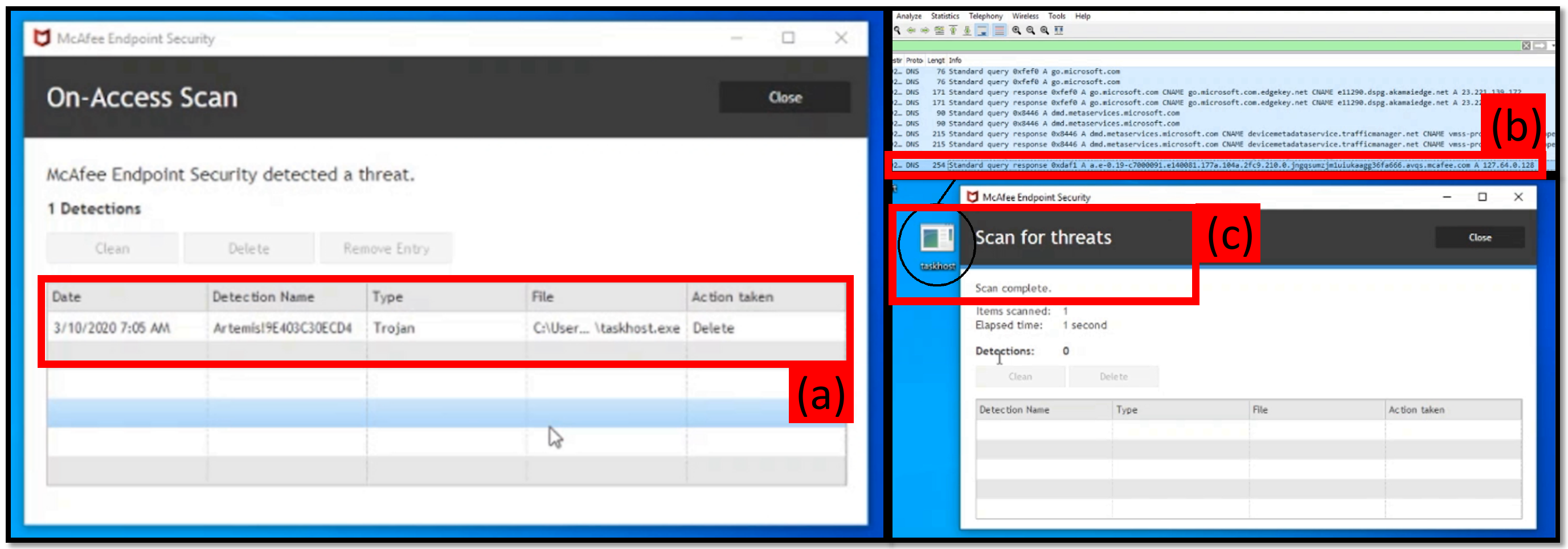}
    \caption{A screenshot the demonstrates the silencing attack on the McAfee ESTP agent.
    The Artemis Trojan, disguised as taskhost.exe, is successfully detected by McAfee ESTP agent when not under attack, and the agent immediately takes the action of deleting it (a).
    The victim downloads the Artemis Trojan and starts ``scan for threats'' using its installed agent.
    A man-in-the-middle attacker spoofs the DNS responses on behalf of the McAfee GTI service to 127.64.0.128 (b).
    Based on the spoofed response, the agent incorrectly classifies the malicious Artemis Trojan as benign, and outputs an empty threat scan report (c).
    The outcome of the attack is that the Artemis Trojan is not detected as a threat and is allowed to be executed on the victim machine.}
    \label{fig:mcafee-silencing-screenshot}
\end{figure*}

We include two screenshots to demonstrate the false alerts attack on Nessus Professional (see Figure~\ref{fig:nessus-fa-screenshot}) and the silencing attack on McAfee ESTP agent (see Figure~\ref{fig:mcafee-silencing-screenshot}).
The screenshots appear on the next page.

\newpage

\balance

\section{\label{sec:responsible_disclosure} Responsible Disclosure}
\textbf{McAfee:} After reaching out to McAfee, they provided the following response: 

"McAfee’s Global Threat Intelligence over DNS (GTI-DNS) service has been in production for about 12 years. At the time of its launch, most internet traffic was not encrypted.  Since then we’ve updated most of our products and cloud services to use encrypted communication facilities such as TLS. The latest evolution of our GTI service is GTI-REST. This runs only over TLS 1.2 and does not support unauthenticated clients. Our GTI-REST POPs receive an A rating from SSL Labs. Our effort to move the remaining core products from GTI-DNS to this new GTI-REST service was already underway when we received the report.
The report identifies that it is possible to snoop DNS traffic and determine GTI’s reputation of a file by examining the A record. An attacker could also gain control of an intermediate DNS server, or they could modify the DNS configuration so as to direct DNS requests to a server under their control. Either of these techniques would enable an attacker to intercept A record queries and send a spoof response that a malicious file is clean.
However, it is important to keep in mind that spoofing a clean A record as described above only has the same effect as when the McAfee product is running on a machine which has no internet connection. Our products use regularly updated local content to ensure that they continue to offer excellent protection when disconnected from the Internet.
Moreover, if an attacker spoofs an A record to suggest to the product that a clean file is malicious, the product verifies this response by requesting a confirmatory TXT record from GTI-DNS. Unlike the A record, the TXT record is digitally signed. This mechanism prevents attackers from orchestrating a DoS by convicting clean files.
We shared this information with Oleg Brodt from Ben-Gurion University and appreciate his decision to look to publish the findings after Q1 2021, allowing us to complete the migration from GTI-DNS to GTI-REST."

\textbf{Tenable:} After we reached out to Tenable with the relevant information, they provided the following response:

"Just as DNS traffic can be used to gain information about which websites are being visited, the DNS traffic used here can be used to infer some information (but not the actual page content). Someone in a position to intercept the DNS traffic of the Nessus scanner or agent could identify that the target is evaluating files against Tenable's hash checking service. While a potential attacker could infer that Tenable considered a file to likely be malicious, they would not be able to identify the file with any certainty, the type of malware, impact on the system, or whether the file may be a false positive.

Overall, our understanding is that the information potentially disclosed as a result of these limitations is difficult to leverage in an attack. While we are not planning to make any immediate changes to the service at this time, we will continue to re-assess the situation for ways to improve in this scenario."

\textbf{Team Cymru:} In response to our responsible disclosure, Team Cymru updated its Community Service documentation and provided the following response: "We have made some changes to our Community Service documentation that highlights the inherent issues with the way unencrypted protocols function.

The issue and scenarios you highlight are not specific and isolated to MHR [Team Cymru’s Malware Hash Registry], but rather with the way the protocol for that delivery method works. We are aware of the issues and make available four delivery methods for exactly that reason.  For people who share your concerns we recommend using the HTTPS interface to the MHR.  The unsigned and unencrypted interfaces to the MHR service are there to support people needing high-rate or low-risk quarrying. Given the massive number of hashes in the MHR, DNSSEC isn’t a practical solution for that zone.  As such, we do not have any plans to alter the availability of functionality of the MHR tool; this is a community service that is hugely popular and that is relied upon by thousands of InfoSec professionals around the world who make the decision as to what their need is and use the appropriate service interface".



\end{document}